# A compact six-beam 3D magneto-optical trap based on multifunctional metasurface chips

Chen Qing[1], Huijie Sun[1], Yize Liu[1], Yuxuan Liao[1], Yongzhuo Li[1,2*], Dengke Zhang[3*], Xue Feng[1,2*], Yidong Huang[1,2*]

[1] Department of Electronic Engineering, Tsinghua University, Beijing 100084, China
[2] Beijing National Research Center for Information Science and Technology, Beijing 100084, China
[3] School of Instrumentation and Optoelectronic Engineering, Beihang University, Beijing 100191, China

*Correspondence: Yongzhuo Li , (liyongzhuo@tsinghua.edu.cn),
Dengke Zhang (dkzhang@buaa.edu.cn),
Xue Feng (x-feng@tsinghua.edu.cn),
Yidong Huang (yidonghuang@tsinghua.edu.cn).

## ABSTRACT

Magneto-optical traps (MOTs), indispensable to cold-atom-based quantum computing and quantum precision measurement, are increasingly desired in compact and stable form. Particularly, the six-beam 3D-MOT, distinguished by symmetric geometry and superior trapping performance, finds particularly broad applications. Conventional implementations, however, rely on complex and bulky free-space optics, hindering miniaturization, practical assembly, and long-term stability. Here, we demonstrate a compact six-beam 3D-MOT system that employs two multifunctional optical metasurface chips to implement the optical-field functionalities required for cooling and trapping atoms. By combining two metasurface chips and a vacuum cell, a photonic-integrated cell module (PICM) is implemented as the core assembly and the volume is only $73 \times 73 \times 107\ \mathrm{mm}^3$. In success, $1.3 \times 10^6$ $^{87}$Rb atoms are trapped and cooled to the temperature of ~183 μK. It should be noticed that the laser beam is guided to the metasurface through optical fiber, thus such fully fiber-coupled architecture enables straightforward and stable operations. We believe that this work would pave a practical route toward compact and portable 3D-MOT.

## INTRODUCTION

Neutral atoms provide an outstanding platform for quantum computing, quantum precision measurement, and quantum sensing owing to the pronounced quantum effects, long coherence time, and high tunability[1–8]. Thus, trapping and cooling atoms are essential for these applications. Among the available techniques, the three-dimensional magneto-optical trap (3D-MOT) is the most widely adopted approach, as it provides a reliable atomic source for subsequent manipulation. The standard configuration employs three pairs of counter-propagating, mutually orthogonal, circularly polarized light beams that intersect at the zero-field point of the quadrupole magnetic field[9–11]. Such a six-beam configuration offers three key advantages: high atom-trapping efficiency, which permits fast loading of high-density cold atoms; a broad optical-access window, which affords enough spatial clearance for the subsequent detection, cooling, and manipulation of cold atoms; and a highly symmetric beam configuration, which substantially simplifies system alignment and practical implementation. Owing to these advantages, it is the most common and mature scheme.

Driven by the increasing demand of integration and miniaturization, compact 3D-MOT schemes are highly desired for various space-constrained applications, such as chip-scale atomic clocks, portable quantum sensors, and satellite-based quantum communication systems[12–16].

Although a six-beam MOT could be well implemented in conventional free-space optics, it is hard to be miniaturized to portable size. Specifically, the three pairs of light beams have to be mutually counter-propagating and orthogonal, while both the polarization and intensity of the beams should be precisely balanced. Thus, it is a significant technical challenge to meet such stringent requirements within a three-dimensional and integrated geometry[17,18]. To address such limitations, some alternative schemes have been proposed, including the pyramidal MOT and the modified structure such as the tetrahedral MOT, as well as the grating MOT[19–31]. In these configurations, four- or five-beam geometries is generated by reflecting or diffracting a single incident beam with mirrors or gratings, combined with wave plates for polarization control. These approaches achieve partially miniaturized MOT system. In particular, the grating MOT offers a compact alternative that significantly simplifies the optical layout, as the diffracted beams are aligned by the grating geometry, and has found applications in quantum measurement. However, the pyramidal MOT would introduce several practical concerns, including diffraction and scattering within the small apex region, additional scattering from the pyramid edges, and absorption-induced shielding of the vertical-axis beam, which would be traversing the atomic cloud twice. Similarly, in the grating MOT, the trapping region locates within a few millimeters upon the chip surface, thereby constraining optical access for subsequent manipulations such as single-atom addressing based on optical tweezers. Moreover, limited by the grating diffraction efficiency and geometry-induced polarization mismatch, the trapping and cooling performance of grating MOT is not as good as that of six-beam configuration.

Recently, nano-photonic devices have opened a promising route toward the miniaturization of six-beam MOT. For example, the photonic-integrated-circuit MOT (PIC-MOT) [32–35] leverages mature photonic integration for high-density on-chip routing and metasurface-based MOT[36,37] utilizes planar optical elements for versatile wave front shaping. However, in PIC-MOT, light beam is guided through waveguides of unequal lengths, which would lead to severely imbalanced output powers among the beams. Furthermore, the low waveguide-to-free-space coupling efficiency further compromises the performance of corresponding quantum applications. Meanwhile, existing free-space metasurface MOT still relies on a considerable number of discrete optical components and the proposed geometric configurations offer only a limited degree of integration.

In this work, we have proposed and demonstrated a compact six-beam 3D-MOT system based on a photonic-integrated cell module (PICM) composed of multifunctional metasurfaces, compact plane mirrors and a regular hexagonal-prism vacuum cell for laser cooling and trapping of $^{87}$Rb atoms. The metasurface can perform polarization conversion, beam deflection, and beam expansion to meet the cooling-beam requirements of a 3D-MOT. The resulting six beams intersect at the magnetic-field zero point, located at the center of the vacuum cell equipped with an anti-Helmholtz coil pair, thereby simultaneously fulfilling the conditions for three-dimensional laser cooling and trapping. In experiment, we present a compact 3D-MOT system that traps and cools more than $10^6$ $^{87}$Rb atoms at the temperature of $\sim 183$ μK. In this system, all input beams are fiber-delivered. The fully packaged device enables straightforward operation, eliminating the tedious alignment and complex adjustment associated with discrete optical elements in conventional systems, thereby greatly enhancing system stability. Moreover, our approach is compatible with high-volume production, paving the way for future large-scale deployment.

## RESULTS

### Design of an integrated six-beam configuration

In a standard 3D-MOT, three mutually orthogonal pairs of counter-propagating, red-detuned cooling beams intersect at a common point, forming an optical molasses. Under a quadrupole magnetic field gradient, the circular handedness of each beam would be properly settled according to the direction of local magnetic field. Specifically, one pair of the light beams would oppositely handed to the other two, thus a position-dependent scattering force could be generated following the Zeeman shift and polarization selection rule[9,10]. Then the atoms would be stored toward the zero-field point as shown in Fig. 1A. Furthermore, all beams require millimeter-scale mode-field diameters (MFD) to maximize the overlap volume and ensure efficient atomic loading.

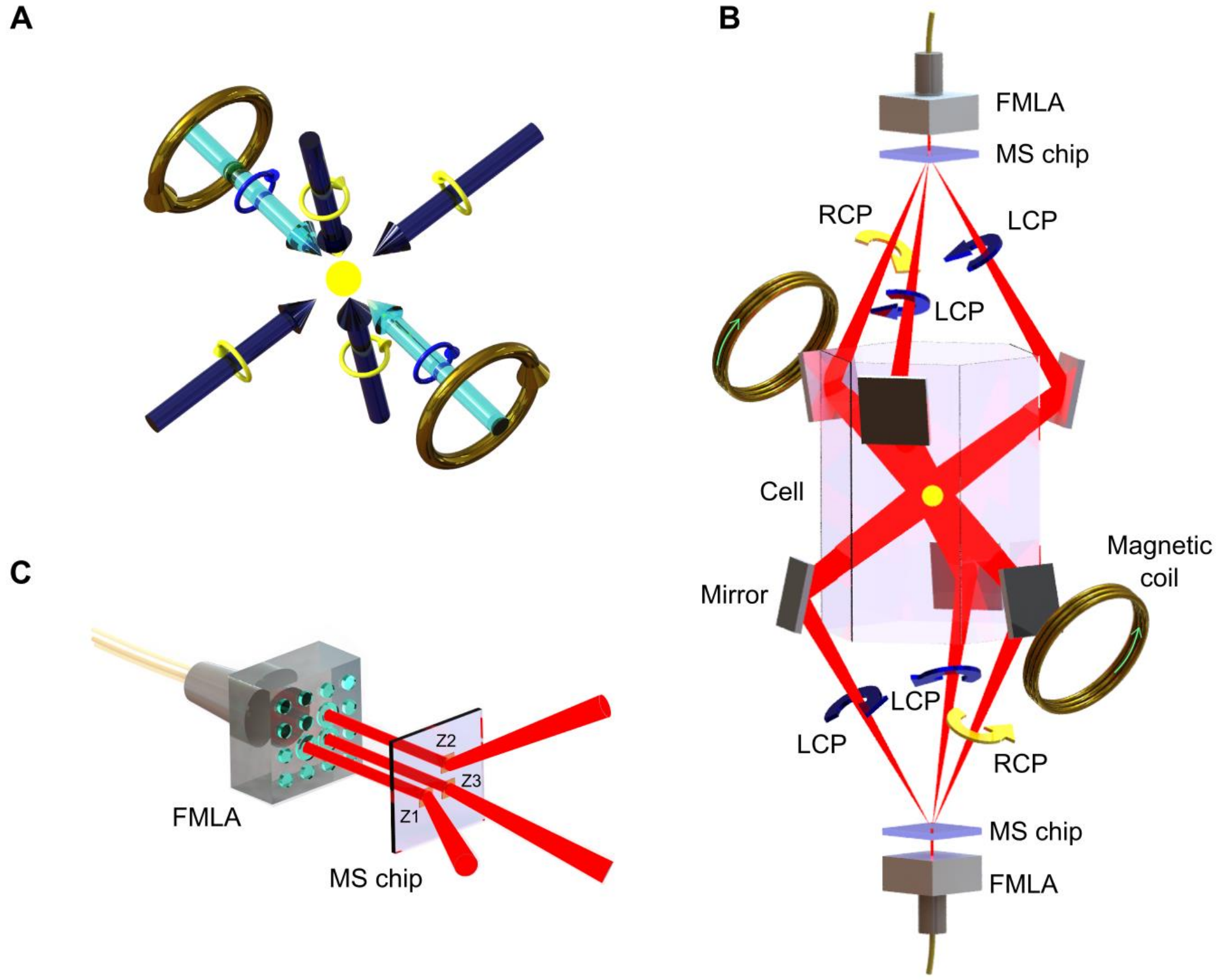

**Figure 1 A** Schematic of the cooling laser beams in a six-beam 3D-MOT. **B** Layout of the core assembly of compact six-beam 3D-MOT system. The system comprises two fiberized micro-lens arrays (FMLAs), two metasurface chips (MS chips), six compact mirrors, a regular hexagonal-prism vacuum cell, and anti-Helmholtz coils. Light from each FMLA illuminates its corresponding metasurface chip, which simultaneously performs polarization conversion, beam deflection, and beam expansion. The six resulting beams are reflected by the mirrors and converge at the cell center, forming three counter-propagating beam pairs in a rotationally symmetric configuration. The anti-Helmholtz coils, aligned along one beam axis, produce a quadrupole magnetic field with a zero at the beam intersection, enabling atom trapping. **C** The FMLA–chip unit comprises three polarization-maintaining input fibers, a lens array, and a metasurface chip. The lens array converts the fiber-guided light into three free-space beams, each aligned with its corresponding nanopatterned zone on the metasurface chip.

To miniaturize the MOT while keeping six-beam configuration, we propose an integrated architecture with optical chips as shown in Fig. 1B. It consists of two identical modules mounted on the top and bottom facets of the vacuum cell. In each module, three incident laser beams, which is delivered by a fiberized micro-lens array (FMLA), would illuminate three independent metasurface zones (Z1, Z2, and Z3) on a single chip as detailed in Fig. 1C. Each zone would perform polarization conversion, beam deflection, and beam expansion, yielding three output beams according to the cooling requirements. Then the three beams are redirected by three compact mirrors and could converge at the cell center. To integrate the chips and mirrors with the vacuum cell tightly and maintain unobstructed access to the cell center over the full horizontal plane, we have employed a regular-hexagonal-prism vacuum cell. For the magnetic field, an anti-Helmholtz coil pair is aligned with one beam pair, whose circular-polarization handedness is opposite to that of the other two, then a quadrupole field could be generated and the center is at the same intersection point. Together, these components form the core architecture of the proposed compact six-beam 3D-MOT.

**Design, fabrication and characterization of the metasurface chip**

The functionalities of the metasurface chip are illustrated in Fig. 2A. Each metasurface zone in the chip transforms a normally incident, linearly polarized Gaussian beam into an output beam deflected by 60° from the chip plane and expanded to the MFD of 5 mm after propagating over 70 mm. The three output beams are azimuthally arranged with a successive 120° rotation according to normal direction of the chip plane. Specifically, Z1 produces right-handed circular polarization (RCP), while Z2 and Z3 produce left-handed circular polarization (LCP) according to the magnetic field arrangement. To achieve the multifunctional manipulations of light-field, we have employed an inverse design strategy: a target phase profile is first derived through numerical optimization, and appropriate meta-atoms are then selected to reproduce the target profile, thereby achieving the final design of the whole chip. Although the three zones share the same design procedure, the phase profile of each zone is individually tailored to satisfy the corresponding polarization and beam-deflection requirements.

Specifically, the metasurface employ rectangular meta-atoms, which is considered as amorphous-silicon (α-Si, $n = 3.65$) patterned on a $SiO_2$ substrate ($n = 1.46$). The high refractive-index contrast between α-Si and $SiO_2$ supports efficient lightwave manipulation at the operation wavelength of 780 nm, which is the according to the Rb atoms. In addition, the birefringence introduced by the rectangular shape enables simultaneous control both the phase and polarization of the transmitted light by engineering the in-plane dimensions (*i.e.*, the length and width) of the meta-atoms. As illustrated in the inset of Fig. 2B, each meta-atom has a height of $h = 500\,\text{nm}$ and is rotated by $45°$ with respect to the lattice axes, forming a square lattice with periods $P_x = P_y = 340$ nm. Finite-difference time-domain (FDTD) simulations are performed to calculate the transmission amplitude and phase delay of meta-atoms while both the length $l$ and width $w$ are within the range from $60\text{ nm}$ to $300\text{ nm}$, thereby constructing a meta-atoms library. Appropriate meta-atoms are then meticulously selected to match the target phase profile (see Supplementary Information for details).

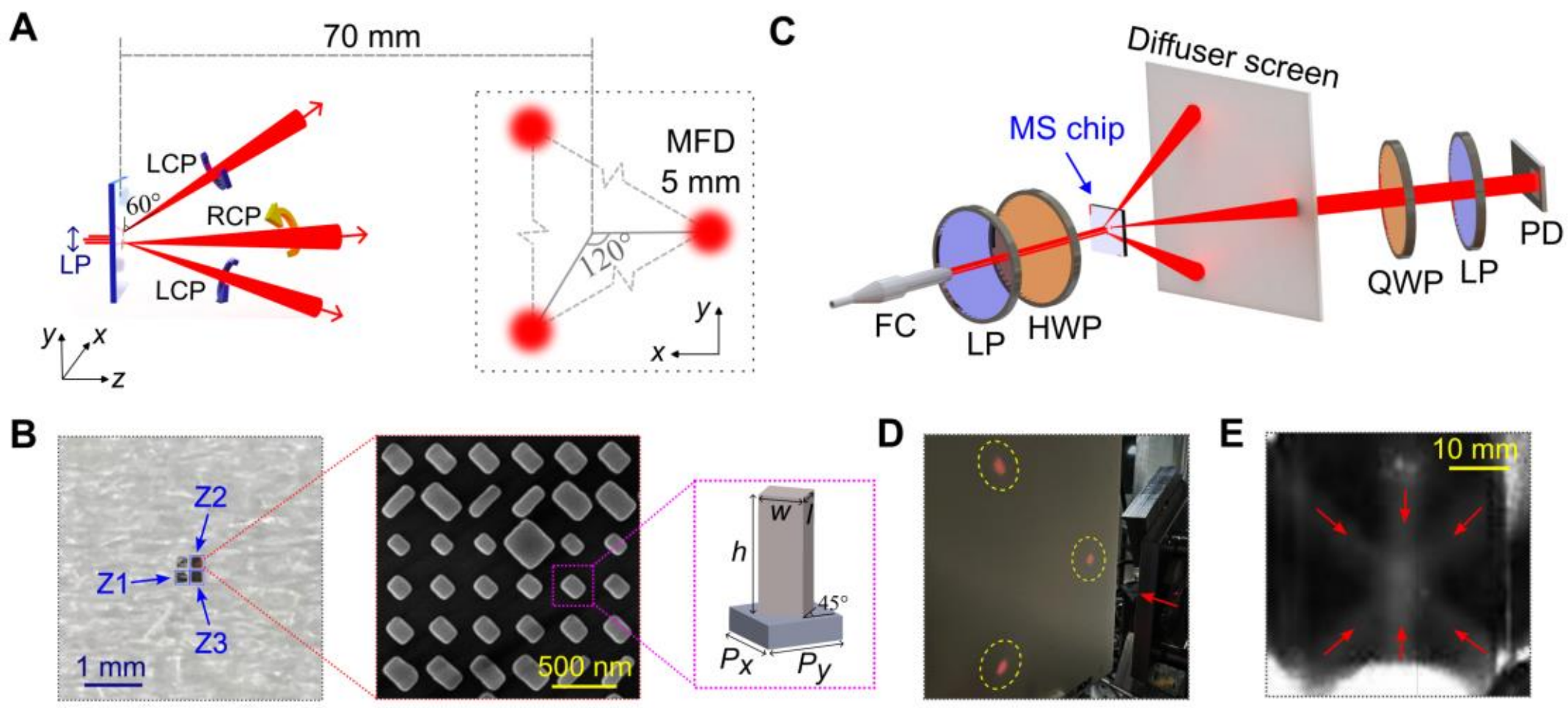

**Figure 1** **A** Schematic of multifunctional optical manipulation by the metasurface chip. Upon normal incidence of linearly polarized light, the output beams are deflected by 60° relative to the surface plane and distributed at 120° intervals in the output plane. The three beams are converted from linear polarization to LCP, LCP, and RCP, respectively, and each is expanded to an MFD of 5 mm after 70 mm of propagation. **B** Characterization of the metasurface chip. From left to right: optical image of the chip, local SEM image, and schematic of the rectangular meta-atom. The chip is $10\,\mathrm{mm} \times 10\,\mathrm{mm}$ , and each metasurface region is $200\,\mu\mathrm{m} \times 200\,\mu\mathrm{m}$. The meta-atom geometry is defined by its periods $\{P_x, P_y\}$, length $l$, width $w$, and height $h$. **C** Schematic of the measurement setup, FC: fiber collimator, LP: linear polarizer, HWP: half-wave plate, QWP: quarter-wave plate, PD: photodetector. **D** Photograph of the three output beam spots, showing the beam deflection and expansion. **E** Six-beam path and the overlap at the cell center.

Based on the aforementioned design, the metasurface chips are fabricated as follows. A 500 nm thick α-Si film is first deposited on a fused silica substrate, and the rectangular meta-atom patterns are then defined by electron-beam lithography (EBL) followed by inductively coupled plasma reactive-ion etching (ICP-RIE). Each metasurface zone occupies an area of $200\,\mu\mathrm{m} \times 200\,\mu\mathrm{m}$, and the size of the entire chip is $10\,\mathrm{mm} \times 10\,\mathrm{mm}$. Figure 2B presents an optical image of the fabricated chip together with a cross-sectional scanning electron microscopy (SEM) image of the meta-atoms.

Before integrated into the MOT system, the chips are first characterized in terms of transmittance, polarization-conversion efficiency, beam deflection, beam expansion, and the overlap of all six beams. the schematic of measurement setup is shown in Fig. 2C. The laser beam from a fiber collimator (FC) passes through a linear polarizer (LP) and a half-wave plate (HWP) and is then normally incident on each of Z1, Z2, and Z3. Thus, the three functional zones can be characterized individually. For each output beam, the transmittance and beam expansion are determined from the measured optical power and the beam profile recorded on a diffuser screen, respectively. Figure 2D shows representative beam images captured on the diffuser screen. The polarization-conversion efficiency (PCE) is further characterized by using a polarization analyzer composed of a quarter-wave plate (QWP) and a LP to extract the LCP and RCP components, respectively. Table 1 summarizes the measured transmittance and polarization conversion efficiencies.

Table 1 Optical measurements of metasurface chips

| MS chip | Zone | Transmittance | PCE |
|---|---|---|---|
| **Sample 1** | Z1 | 20.9% | 88.5% |
| | Z2 | 21.4% | 89.6% |
| | Z3 | 20.2% | 89.4% |
| **Sample 2** | Z1 | 34.2% | 93.8% |
| | Z2 | 30.8% | 93.4% |
| | Z3 | 31.6% | 91.6% |

For sample 1, the measured transmittances of Z1, Z2, and Z3 are 20.9%, 21.4%, and 20.2%, respectively, while the corresponding PCE are 88.5%, 89.6%, and 89.4%. For sample 2, the measured transmittances are 34.2%, 30.8%, and 31.6%, respectively, with corresponding PCE of 93.8%, 93.4%, and 91.6%. Since the available laser power is about 2 W, the measured transmittance is sufficient to meet the requirements of a 3D-MOT. Owing to fabrication variations between the two batches, there is some difference between the overall performance of the two samples. Nevertheless, the output powers of the three metasurface zones within each chip remain well balanced. Therefore, the deviations between two chips can be compensated by adjusting the input power. Finally, the intersection of the six beams inside the Rb vapor cell is verified by fluorescence imaging, in which two metasurface chips are combined with six compact mirrors. Figure 2E shows the propagation beam paths of the six beams and the corresponding intersection at the cell center. These results indicate that the fabricated chip samples agree well with the design and could meet the requirement of manipulating the light-field in a 3D-MOT.

**Assembly and characterization of the compact 3D-MOT**

In success, a MOT system has been implemented and measured as shown in Figure 3A, in which there are two independent lasers for cooling and repump, respectively. The cooling laser is frequency-stabilized by saturation absorption spectroscopy (SAS) and shifted by an acousto-optic modulator (AOM) to a red detuning of approximately $13.3\ \mathrm{MHz}$ from the $F_{\mathrm{g}} = 2 \rightarrow F_{\mathrm{e}} = 3$ hyperfine transition of the $^{87}$Rb D2 line. The repump laser is resonant with the $F_{\mathrm{g}} = 1 \rightarrow F_{\mathrm{e}} = 2$ transition of the $^{87}$Rb D2 line. The cooling and repump light beams are combined by a fiber coupler at a $75:25$ ratio and then equally split into two copies, each of which would pass through a $1 \times 3$ fiber splitter and yield collimated light beams into the FMLA.

The FMLAs deliver light beams to the metasurface chips, which are positioned symmetrically above and below the vacuum cell. Each FMLA–chip unit is secured by a mechanical fixture and mounted facing the corresponding cell facet. The vertical position of each unit is adjustable, enabling fine tuning of the six-beam intersection point inside the cell. Each beam out from the metasurface is then redirected by a compact mirror (10 mm × 10 mm square, 1 mm thick), which is placed along the corresponding output beam path. The mirror is settled with tilted angle of 12° corresponding to the cell sidewall and positioned at a 5 mm gap so that all six beams would enter the vacuum cell with an incidence angle of 36° and converge at the center of the cell. The regular hexagonal-prism vacuum cell has a height of 40 mm, a face-to-face distance of 28 mm, and a wall thickness of 2 mm. It is evacuated to a background pressure of $5\times10^{-8}$ Pa, achieving a long-term stable vacuum level better than $1\times10^{-6}$ Pa following atom release. The footprint of the PICM is 73 × 73 × 107 mm$^3$, as shown in the inset of Fig. 3A. An anti-Helmholtz coil pair placed on either side of the PICM produces a quadrupole magnetic field with a gradient of ~15 G/cm along the coil axis at the cell center. This configuration ensures that the six beams converge at the zero point of the quadrupole magnetic field at the cell center.

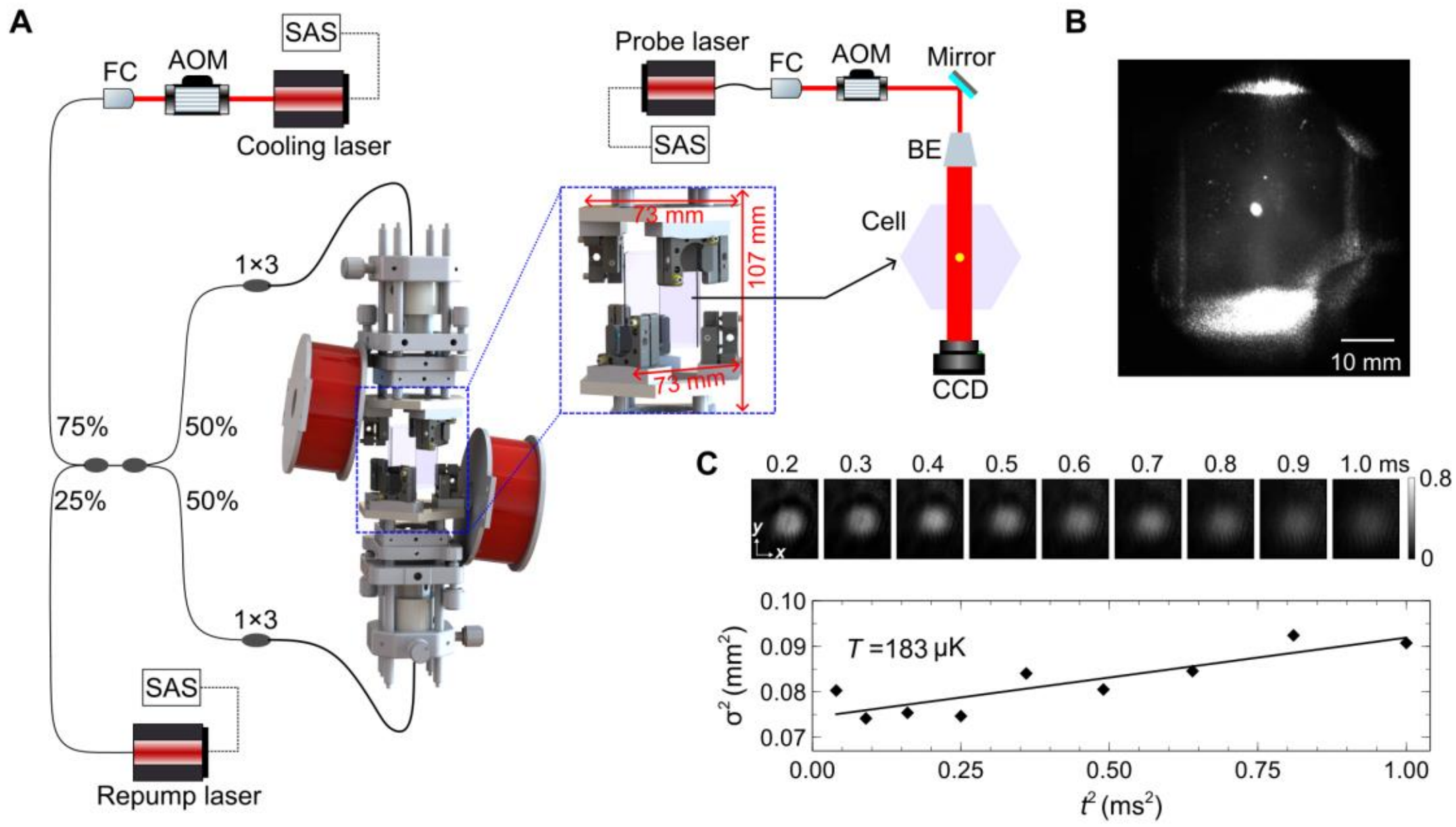

**Figure3** Performance of cold atoms trapped in a compact 3D-MOT. **A** Schematic of the complete compact 3D-MOT system. Left: laser cooling and trapping system. Right: absorption-imaging-based detection system. Insert: magnified view and dimensions of the PICM. SAS: saturation absorption spectroscopy, AOM: acousto-optic modulator, BE: beam expander, CCD: charge-coupled device. **B** Fluorescence image of the trapped cold atomic cloud. **C** Cold atomic clouds at different times of flight (from 0.2 ms to 1 ms in steps of **0.1** ms) and the corresponding temperature fitting curve.

In the experiment, each cooling beam and repump beam are delivered with an optical power of 2 mW and 0.05 mW, respectively. The fluorescence images of the cold atoms are captured by a charge-coupled device (CCD) camera, as shown in Fig. 3B. To determine the number and temperature of trapped atoms, an absorption-imaging-based detection system has been employed, as shown in the right inset of Fig. 3A. The probe beam is locked to the $F_g = 2 \rightarrow F_e = 3$ transition of the $^{87}$Rb D2 line. By switching off the cooling lasers, the atoms would expand freely, and absorption images are captured with a CCD camera at various times of flight. The atom cloud size is then extracted by fitting the optical depth (OD) as: $\mathrm{OD} = -\ln(I_{\mathrm{atom}}/I_{\mathrm{ref}})$, where $I_{\mathrm{atom}}$ denotes the transmitted probe light intensity detected by the CCD in the presence of trapped atoms, and $I_{\mathrm{ref}}$ denotes the referenced light intensity detected in the absence of atoms. Figure 3C presents OD images of the atom cloud acquired at time-of-flight (TOF) values ranging from 0.2 ms to 1 ms in steps of 0.1 ms. The atom number $N$ is obtained by summing the OD over the region of interest, multiplying by the object-plane pixel area ($5.86\ \mu\mathrm{m} \times 5.86\ \mu\mathrm{m}$), and divided by the resonant absorption cross-section $\sigma_0 = \lambda^2/2\pi$. The number of trapped atoms is extracted from all TOF images and then averaged, yielding $N = 1.3 \times 10^6$. Meanwhile, the mean cloud size $\sigma^2$ is obtained from the OD by averaging the variances of the atomic cloud along the $x$ and $y$ directions. The cloud size $\sigma^2$ at different expansion times is then linearly fitted as a function of $t^2$ according to $\sigma^2 = \sigma_0^2 + \left(\frac{k_B T}{m}\right) t^2$, here $k_B$ is the Boltzmann constant and $m$ is the mass of a single $^{87}$Rb atom, from which the mean temperature $T = 183\ \mu\mathrm{K}$ is thus extracted.

## DISCUSSION

In summary, we have demonstrated a compact six-beam 3D-MOT, in which a core PICM volume is 73 × 73 × 107 $mm^3$ while more than $10^6$ cold atoms are trapped and cooled at a temperature of approximately 183 μK. Central of this miniaturized architecture is a multifunctional metasurface chip that monolithically integrates polarization conversion, beam deflection, and beam expansion, thereby fulfilling all beam-manipulation functions required for cooling and trapping atoms. By eliminating the cascaded waveplates and associated optical elements in conventional systems, this metasurface-based approach significantly reduces system complexity and advances the miniaturization and integration of MOT systems. Furthermore, the proposed integration scheme establishes the feasibility of all-optical miniaturization for six-beam 3D-MOT systems, paving the way for portable cold-atom instruments in volume-constrained environments, such as satellite payloads and vehicle-mounted platforms.

In the current implementation, owing to differences of fabricating two chip samples, the refractive indices of *α*-Si differ slightly, resulting in some discrepancies in the optical performance. Nevertheless, according to the practical requirements of the MOT system and the final experimental results, such deviation does not introduce notable impact on the overall performance. Furthermore, the compatibility of our approach with established nanofabrication processes renders it amenable to large-volume production. In future implementations, vertically emitting photonic chips with integrated lasers could consolidate light generation, frequency shifting, and beam splitting onto a single device. Combined with the metasurface in a stacked two-layer architecture, this strategy would enable monolithic integration of all optical functions required for MOT systems, further reducing the system footprint and ultimately achieving a compact, lightweight, and low-power integrated MOT system. Even in its present form, the demonstrated MOT already exhibits competitive performance while the achieved temperature is very close to the Doppler limit (183 μK vs. 146 μK). With this result, a two-dimensional MOT could also be introduced as a pre-cooling stage to further increase the trapped-atom number, followed by polarization gradient cooling to push the temperature below the Doppler limit. Such issue is still under investigation.

### Fabrication process

The metasurface chips used in this work were fabricated on fused silica substrates with an amorphous silicon (α-Si) functional layer. First, a 500 nm thick α-Si film was deposited on a 500 μm thick fused silica substrate by plasma-enhanced chemical vapor deposition (PECVD). A 170 nm thick layer of ZEP520A positive-tone electron-beam resist was then spin-coated onto the α-Si surface. After soft baking, electron-beam lithography (EBL) was performed with a beam current of 2 nA to transfer the meta-atom patterns into the resist layer. The exposed sample was subsequently immersed in developer for 90 s and in fixer for 30 s to form the resist mask. Next, a 50 nm thick chromium (Cr) layer was deposited by electron-beam evaporation. The sample was then immersed in hot acetone to lift off the resist and the overlying Cr, leaving Cr patterns directly on the α-Si surface and thereby forming a Cr hard mask that transferred the pattern from the resist layer to the α-Si layer. Subsequently, the α-Si layer was anisotropically etched by inductively coupled plasma reactive-ion etching (ICP-RIE), transferring the pattern into the α-Si layer. Finally, the residual Cr mask was removed by wet etching in a Cr etchant, yielding the designed metasurface pattern in the α-Si layer. A schematic of the fabrication process is provided in Supplementary S3.

**MOT Implementation**

In the compact MOT system, the cooling and repump beams were derived from distributed Bragg reflector (DBR) lasers with a center wavelength of 780 nm. Using saturated absorption spectroscopy, the cooling laser was locked to the $5^2\mathbf{S}_{1/2}\,(F_\mathrm{g}=2)\rightarrow 5^2\mathbf{P}_{3/2}\,(F_\mathrm{e}=2{,}3)$ crossover transition, and an acousto-optic modulator (AOM) was used to shift its frequency by approximately 120 MHz, yielding a red detuning of about 13.3 MHz from the $5^2\mathbf{S}_{1/2}\,(F_\mathrm{g}=2)\rightarrow 5^2\mathbf{P}_{3/2}\,(F_\mathrm{e}=3)$ cooling transition. Similarly, the repump laser was locked to the $5^2\mathbf{S}_{1/2}\,(F_\mathrm{g}=1)\rightarrow 5^2\mathbf{P}_{3/2}\,(F_\mathrm{e}=2)$ transition by saturated absorption spectroscopy. The frequency-stabilized beams were then coupled into optical fibers and, after fiber-based combining and splitting, delivered to a fiber lens array. The lens array was mounted in an aluminum alloy holder, while the metasurface chip was fixed in a 3D-printed plastic mount. The chip and the fiber lens array were integrated into a cage system, with the chip positioned approximately 8 mm from the output facet of the lens array. The mirrors were mounted within the same cage structure, enabling a fully integrated assembly. In the experiment, the height of a translation stage was adjusted to position the vacuum cell within the six-mirror assembly, completing the compact MOT setup. Further details are provided in Supplementary S4. Atom number and temperature were measured using an independent probe beam path. The probe beam passed through the atomic cloud and was imaged onto a CCD camera. By analyzing the acquired absorption images, the atom number and temperature were extracted.

**ACKNOWLEDGEMENTS**

This research was supported by the funding from Beijing National Research Center for Information Science and Technology (BNRist, BNR2026ZD01002), the National Natural Science Foundation of China (Grant No. 92576208) is greatly acknowledged. This work was also supported by TSINGXIN Quantum Corporation, the Frontier Science Center for Quantum Information, the Beijing Academy of Quantum Information Sciences, and the Tsinghua University Initiative Scientific Research Program. The processes of fabricating the samples are technically supported by the TIANJIN H-CHIP TECHNOLOGY GROUP CORPORATION.

**CONFLICT OF INTEREST**

The authors declare no conflict of interest.

**AUTHOR CONTRIBUTIONS**

Chen Qing, Dengke Zhang and Xue Feng conceived the idea. Chen Qing, Dengke Zhang, Huijie Sun and Yize Liu designed and performed the simulations, experiments, and data analysis. Yuxuan Liao contributed significantly to the phase pattern optimization algorithm based on gradient descent. Yongzhuo Li provided useful discussions and comments. Chen Qing, Dengke Zhang and Xue Feng wrote the paper. Chen Qing, Dengke Zhang, Xue Feng, Yongzhuo Li and Yidong Huang revised the manuscript. Xue Feng and Yidong Huang supervised the research. The manuscript was written through contributions of all authors. All authors have given approval to the final version of the manuscript.

**DATA AVAILABILITY STATEMENT**

The data that support the findings of this study are available from the corresponding authors upon reasonable request.

Supporting Information for:

# A compact six-beam 3D magneto-optical trap based on multifunctional metasurface chips

## S1. Beam propagation and assembly design of the compact 3D MOT

The three light beams from the fiberized micro-lens array (FMLA) are directed onto three distinct patterned zones (Z1-Z3) of the metasurface chip, which simultaneously perform polarization conversion, beam deflection, and beam expansion, as illustrated by the three output beams in Fig. S1A. Specifically, the left zone (Z1) converts the incident beam into right-handed circularly polarized (RCP) light and deflects it left at an angle of 60° relative to the chip plane, while the two right zones (Z2, Z3) convert the corresponding incident beams into left-handed circularly polarized (LCP) light and deflect the beams toward the upper-right and lower-right directions at the same 60°, respectively. After propagating over a distance of 70 mm, such three light beams would expand to a diameter of 5 mm and are azimuthally separated by 120° in the transverse plane.

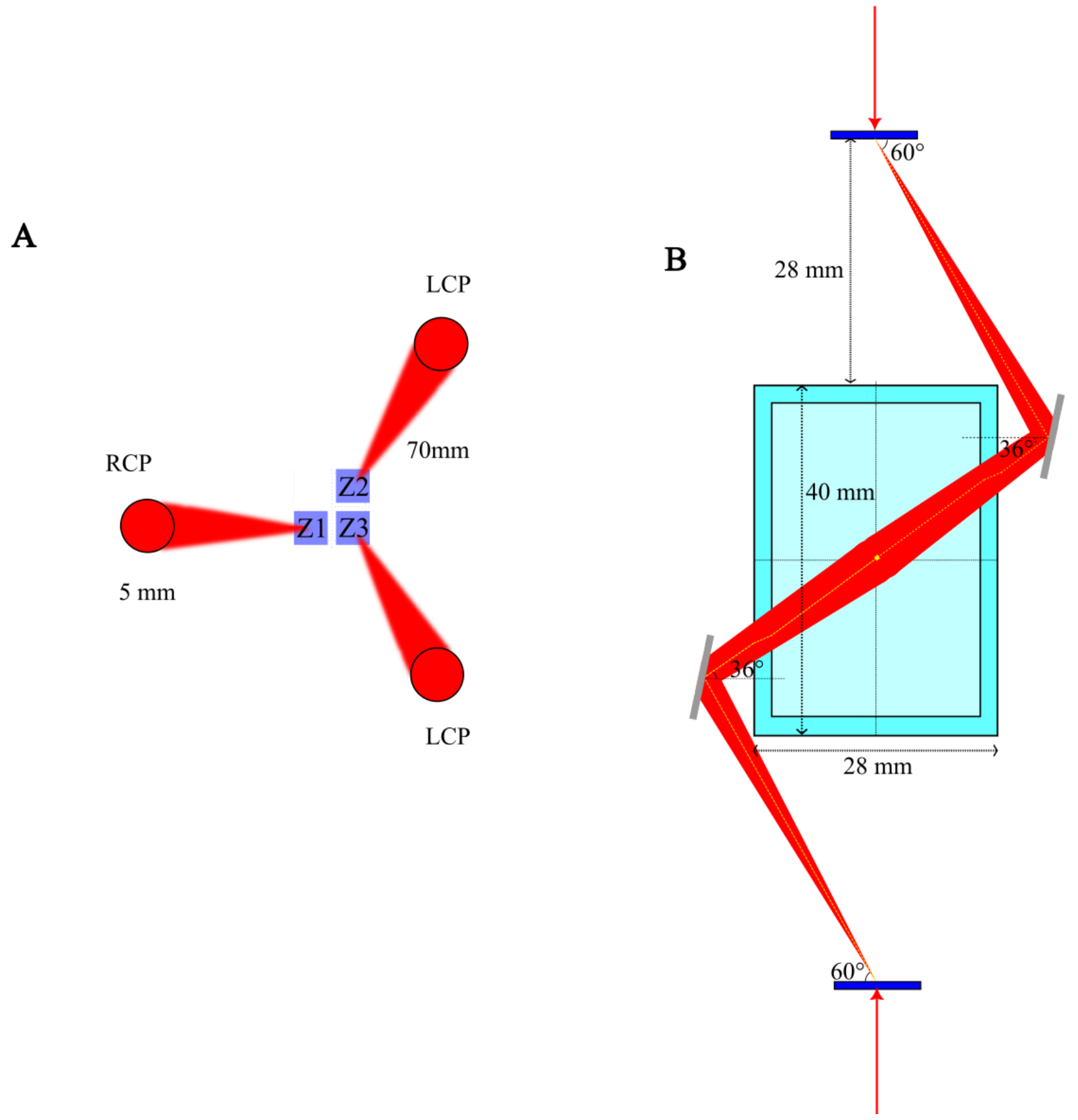

**Figure S1 A** Output beam distribution after the metasurface chip. **B** Schematic illustration of the beam path for a pair of beams propagating through the metasurface, mirrors, and vacuum cell.

In this work, we employ a regular hexagonal-prism vacuum cell with a height of 40 mm and an opposite-side distance of 28 mm. Two metasurface chips are positioned at the top and bottom ends of the cell, respectively. Fig. S1B illustrates the beam path for one of the three beam pairs. To accommodate beam expansion during propagation, the chips are placed 28 mm away from the top (or bottom) face of the vacuum cell. This spacing prevents the beams from clipping the cell walls, which would otherwise introduce stray reflections, while keeping the overall package compact. The mirrors are tilted by 12°, so that the reflected beams enter the vacuum cell at an angle of approximately 36° relative to its end faces. The upper and lower beams counter-propagate and overlap at the center of the cell. Extending this configuration to the remaining two beam pairs yields three pairs of mutually orthogonal beams that converge at the cell center, thereby satisfying the optical configuration required for the MOT.

## S2. Simulation and design of the metasurface for the compact MOT

The metasurface chip is designed through the following procedure. Rectangular α-Si meta-atoms, rotated by 45° relative to the lattice axes, are adopted in this work. First, a meta-atom library is established by calculating the phase responses $\varphi_x$ and $\varphi_y$ for *x*- and *y*-polarized light under normal incidence, together with the corresponding transmission magnitudes $A_x$ and $A_y$, for rectangular meta-atoms with varying lengths and widths. This library then guides the stepwise design of the metasurface toward the targeted multifunctional performance. To achieve linear-to-circular polarization conversion, each meta-atom is required to serve as a quarter-wave plate, which imposes two conditions: the phase-difference condition $\Delta\varphi = \varphi_x - \varphi_y = 90^{\circ}$ or $270^{\circ}$, and the equal-amplitude condition of $A_x/A_y \approx 1$. The geometric parameters of the meta-atoms fulfilling both conditions are collected to form a candidate library for polarization conversion. Figures S2A and S2C present the distributions of $\Delta\varphi$ and $A_x/A_y$, respectively, where the red (blue) circles mark the meta-atom geometries satisfying the 90° and 270° phase-difference condition together with $A_x/A_y \approx 1$.

Building on this foundation, beam deflection and expansion are further implemented. Such two functionalities are governed by the propagation phase $\varphi_x$, the distribution of which is shown in Fig. S2B. Mapping the meta-atoms preselected for polarization conversion (i.e., the points marked by red and blue circles in Fig. S2) onto the $\varphi_x$ distribution reveals that these points span the full 0–2π phase shift, enabling the construction of arbitrary phase modulation profiles. In addition, to preserve the performance of the metasurface chip, the selected meta-atoms are required to exhibit only minor variations in transmittance $T = A_x^2 + A_y^2$ between adjacent elements. Accordingly, as shown in Fig. S2D, structures with high transmittances and the variation of less than 10% among the mapped points are chosen. In this way, a final meta-atom library is established that simultaneously satisfies the phase condition for linear-to-circular polarization conversion while enabling beam deflection and expansion.

Subsequently, a discrete Fourier transform is employed to generate a phase profile that satisfies the prescribed deflection angle and beam-expansion requirements, and a gradient-descent algorithm iteratively optimizes this profile toward the ideal performance target. The optimized phase profile is then mapped onto the corresponding meta-atom geometries in the library. Finally, the phase profile and the meta-atom arrangement are jointly configured to achieve linear-to-circular polarization conversion under *x*-polarized incidence. In this way, the design of a monolithic metasurface device integrating polarization conversion, beam deflection, and beam expansion is completed.

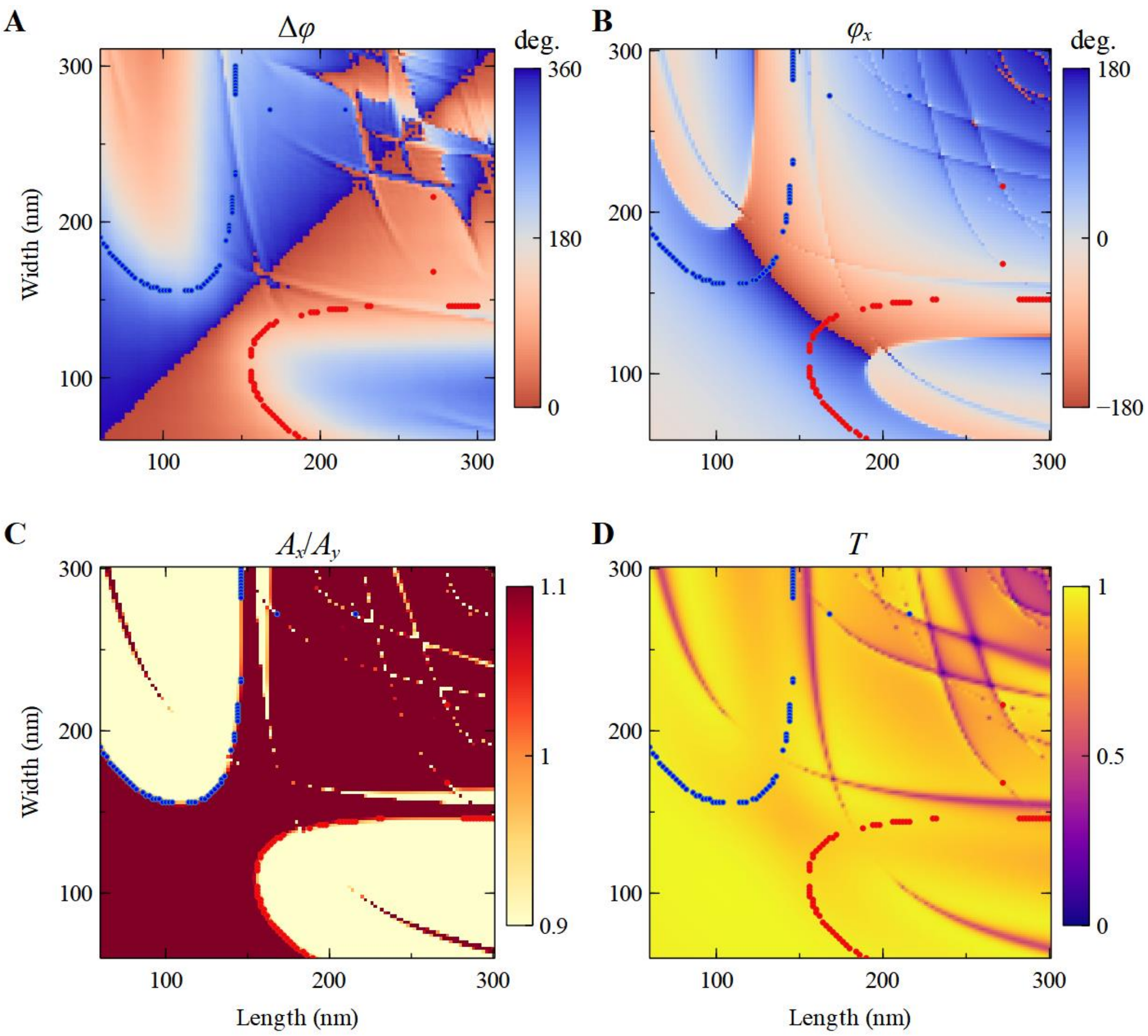

**Figure S2** Basic optical features of uniform metasurfaces. **A** Phase difference distribution $\Delta\boldsymbol{\varphi}$ between $\boldsymbol{\varphi_x}$ **and** $\boldsymbol{\varphi_y}$. **B** Propagation phase distribution $\boldsymbol{\varphi_x}$. **C** Amplitude condition $\boldsymbol{A_x/A_y}$. **D** Transmittance *T*.

## S3. The nanofabrication of the metasurface chip

The process of fabricating the metasurface chip is illustrated in Fig. S3. A 500-nm-thick amorphous silicon ($\alpha$-Si) film was deposited onto a 500-μm-thick fused silica substrate via plasma-enhanced chemical vapor deposition (PECVD). A positive-tone electron-beam resist (ZEP520A) is then spin-coated onto the a-Si film to a thickness of approximately 170 nm. As a positive-tone resist, the exposed regions are selectively removed upon development. After spin coating, the resist is baked and subsequently patterned by electron-beam lithography (EBL) at a beam current of 2 nA, thereby transferring the designed meta-atom patterns into the resist layer with high fidelity. The exposed sample was then immersed in a developer for 30 s and in a fixer for 30 s to form the resist mask structure.

Afterwards, a 50-nm-thick chromium (Cr) layer is deposited onto the resist mask by electron-beam evaporation. The sample is then immersed in hot acetone to lift off the resist together with the overlying Cr, leaving behind a Cr hard mask and thereby transferring the pattern from the resist layer to the Cr layer. Subsequently, the exposed amorphous silicon film is etched by inductively coupled plasma reactive-ion etching (ICP-RIE), further transferring the meta-atom pattern into the silicon film. Finally, the chip is immersed in a Cr etchant to remove the residual Cr mask from the silicon surface, yielding the designed amorphous silicon metasurface chip.

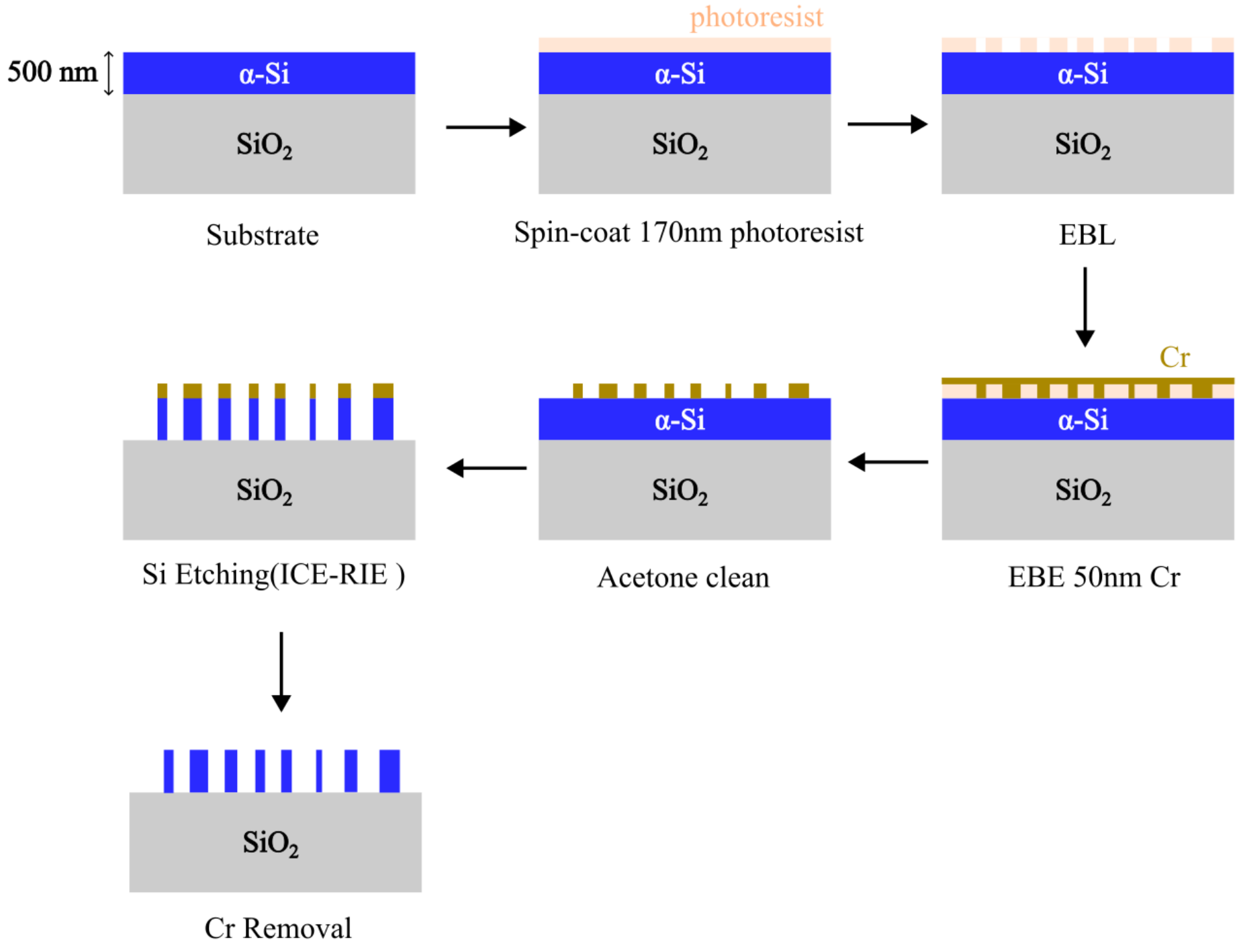

**Figure S3** The fabrication process of the metasurface chip.

## S4. Core component assembly of the compact MOT

The core components of the compact MOT include the metasurface-based beam-manipulation optics, the vacuum cell, and the magnetic-field coils. To facilitate alignment and experimental handling, the optical assembly is built on a cage system, in which the FMLA-chip-unit holder, and the mirrors are integrated into a single module mounted on a three-axis translation stage. By adjusting the three-axis stage, the position of the cage assembly is tuned to coincide with the magnetic-field center while maintaining beam-pointing stability. The optical system is divided into upper and lower modules, such that the convergence point of the six beams can be adjusted vertically through the stage translation. Fig. S4 shows the convergence of the three beams in the upper module, demonstrating that proper adjustment of the stage height brings these beams into focus. Final co-alignment of the upper and lower modules brings all six beams into convergence at the center of the vacuum cell.

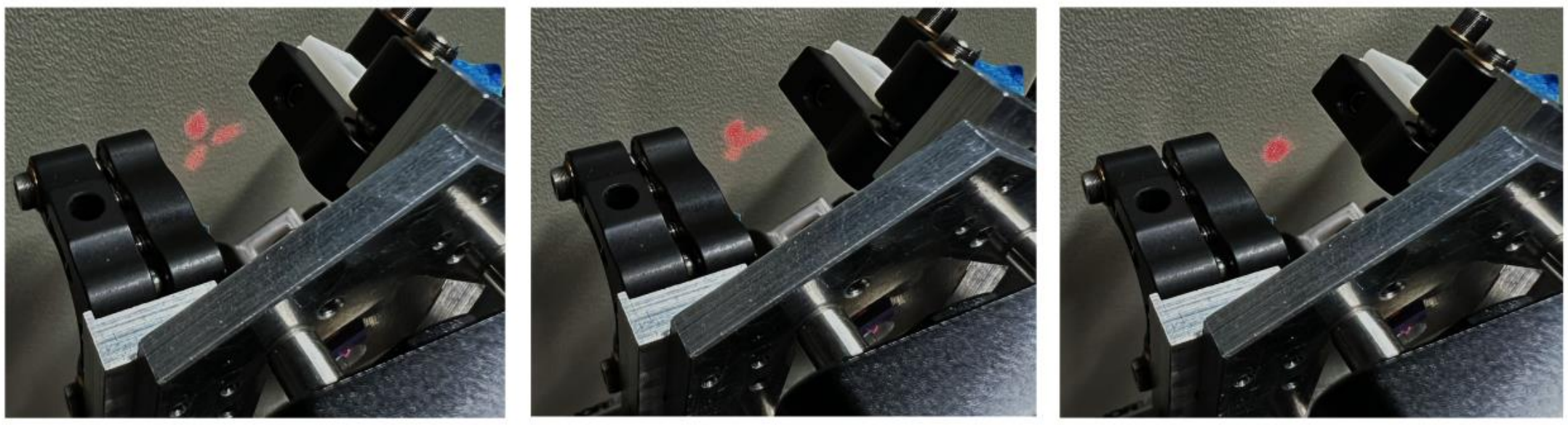

**Figure S4** Single-sided optical path in the core region of the compact 3D-MOT, showing the gradual convergence of the three beams as the imaging plane moves from far to near.

The anti-Helmholtz coils are then positioned on either side of the beam path whose polarization handedness differs from that of the other two paths, with the axis connecting the two coil centers aligned along this beam path. This configuration yields a zero magnetic field at the trap center, a uniform field distribution over a central volume of 1 cm³, and a field gradient of 15 G/cm along the coil axis. Based on this arrangement, the experimental core of the compact MOT is constructed, as illustrated in Fig. S5. For engineering implementation, the system can be further streamlined by integrating the FMLA, the metasurface chip, and the three mirrors assemblies into a single module. Two identical modules are then mounted above and below the vapor cell, followed by the installation of the magnetic-field coils, thereby completing the engineered compact-MOT core.

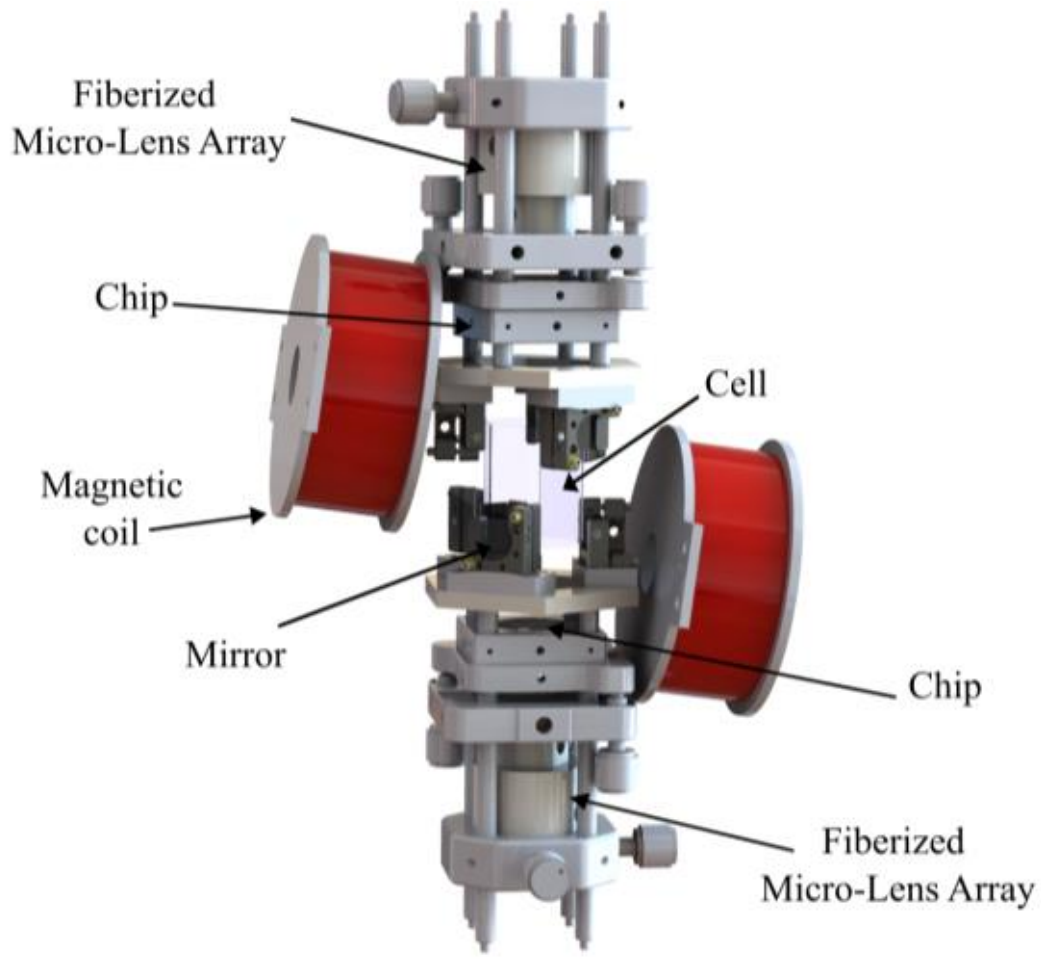

**Figure S5** Assembly model of the compact 3D-MOT core region.